\documentclass[aoas,MSNbibl,nameyear,dvips]{arximspdf}
\usepackage{graphicx}
\usepackage{breakurl}

\doi{10.1214/11-AOAS464}
\volume{5}
\issue{3}
\pubyear{2011}
\firstpage{1699}
\lastpage{1725}

\makeatletter

\def\pr{\operatorname{pr}}
\def\TIC{\mathrm{TIC}}
\def\CLIC{\mathrm{CLIC}}
\newcommand{\eqref}[1]{(\ref{#1})}

\makeatother

\begin{document}
\begin{frontmatter}

\title{Spatial modeling of extreme snow depth}
\runtitle{Spatial modeling of extreme snow depth}

\begin{aug}
\author[A]{\fnms{Juliette} \snm{Blanchet}\ead[label=e1]{Juliette.Blanchet@epfl.ch}}
and
\author[A]{\fnms{Anthony C.} \snm{Davison}\corref{}\ead[label=e2]{Anthony.Davison@epfl.ch}\thanksref{t1}}
\thankstext{t1}{Supported in part by the Swiss FNS and the ETH domain
Competence Center Environment and Sustainability project EXTREMES
(\protect\url{http://www.cces.ethz.ch/projects/hazri/EXTREMES}).}
\runauthor{J. Blanchet and A. C. Davison}
\affiliation{Ecole Polytechnique F\'{e}d\'{e}rale de Lausanne}
\address[A]{Ecole Polytechnique F\'{e}d\'{e}rale de Lausanne\\
EPFL-FSB-MATHAA-STAT \\
Station 8, 1015 Lausanne\\
Switzerland\\
\printead{e1}\\
\hphantom{E-mail: }\printead*{e2}} %adresu isvedimo komanda gale!
\end{aug}

% HISTORY:
\received{\smonth{7} \syear{2010}}
\revised{\smonth{2} \syear{2011}}

% ABSTRACT
%
\begin{abstract}
The spatial modeling of extreme snow is important for adequate risk
management in Alpine and high altitude countries. A natural approach
to such modeling is through the theory of max-stable processes, an
infinite-dimensional extension of multivariate extreme value theory. In
this paper we describe the application of such processes in modeling
the spatial dependence of extreme snow depth in Switzerland, based on
data for the winters 1966--2008 at 101 stations. The models we propose
rely on a climate transformation that allows us to account for the
presence of climate regions and for directional effects, resulting
from synoptic weather patterns. Estimation is performed through
pairwise likelihood inference and the models are compared using
penalized likelihood criteria. The max-stable models provide a much
better fit to the joint behavior of the extremes than do independence
or full dependence models.
\end{abstract}

% KEYWORDS
%

\end{frontmatter}

%s1 ###
\section{Introduction}

Heavy snow events are among the most severe natural hazards in
mountainous countries. Every year, winter storms can hinder mobility by
disrupting rail, road and air traffic. Extreme snowfall can overload
buildings and cause them to collapse, and can lead to flooding due to
subsequent melting. Deep snow, combined with strong winds and unstable
snowpack, contributes to the formation of avalanches, and can cause
fatalities and economic loss due to property damage or reduced
mobility. The quantitative analysis of extreme snow events is important
for the dimensioning of avalanche defence structures, bridges and
buildings, for flood protection measures and for integral risk management.

Compared to phenomena such as rain, wind or temperature, extreme-value
statistics of snow has been little studied. \citet{bocchiola06} and
\citet{bocchiola08} analyzed three-day\vadjust{\eject} snowfall depth in the
Italian
and Swiss Alps, and more recently \citet{blanchet09} analyzed extreme
snowfall in Switzerland. These articles derive characteristics of
extreme snow events based on univariate extreme-value modeling which
does not account for the dependence across different stations. The
spatial dependence of extreme snow data has yet to be discussed in the
literature.

Statistical modeling with multivariate extreme value distributions
began around two decades ago with publications such as \citet{tawn88}
and \citet{coles91}, and has subsequently often been used for
quantifying extremal dependence in applications. Financial examples are
currency exchange rate data [\citet{hauksson01}], swap rate data
[\citet{hsing04}] and stock market returns [Poon, Rockinger and
Tawn (\citeyear{PoonRockingerTawn2003,PoonRockingerTawn2004})], and environmental
examples are rainfall data [Schlat\-her and Tawn (\citeyear{schlather03})], oceanographical data
[\citet{dehaan98}; \citet{coles94}] and wind speed data [\citet
{ColesWalshaw1994}; \citet{FawcettWalshaw2006}]. None of these articles treats
the process under study as a spatial extension of multivariate extreme
value theory.

Until recently, a key difficulty in studying extreme events of spatial
processes has been the lack of flexible models and appropriate
inferential tools. Two different approaches to overcome this have been
proposed. The first and most popular is to introduce a latent process,
conditional on which standard extreme models are applied
[\citet{coles98}; \citet{FawcettWalshaw2006}; \citet{cooley06}; \citet{cooley07};
\citet{gaetan07}; \citet{sang08}; \citet{eastoe08}].
Such models can be fitted using Markov chain Monte Carlo simulation,
but they postulate independence of extremes conditional on the latent
process, and this is implausible in applications. One approach to
introducing dependence is through a spatial copula, as suggested by
\citet{SangGelfand2009Continuous}, but although this approach is an
improvement, \citet{DavisonPadoanRibatet2010} show that it can
nevertheless lead to inadequate modeling of extreme rainfall. A second
approach now receiving increasing attention rests on max-stable
processes, first suggested by \citet{dehaan84} and developed by, for
example, \citet{schlather02} and \citet{kabluchko09}.
Recent applications
to rainfall data can be found in \citet{BuishanddeHaanZhou2008}, \citet
{smith09}, \citet{padoan10} and \citet
{DavisonPadoanRibatet2010}, and to
temperature data in \citet{davison10}. Max-stable modeling has the
potential advantage of accounting for spatial dependence of extremes in
a way that is consistent with the classical extreme-value theory, but
is much less well developed than the use of latent processes or
copulas.
%: five stations in \citet{smith09}; $46$ stations over an area
%around
%ten times the area of Switzerland in \citet{padoan10}; and $14$
%stations over Switzerland in \citet{davison10}.\footnote{Modify
%this?}
%%They possibly use too simple data---and therefore too simplistic
%models--in regard to the real variability of climate data.

In the present paper, we use data from a denser measurement network
than for previous applications. Owing to complex topography and weather
patterns, the processes of \citet{schlather02} and \citet
{smith91} cannot
account for the joint distribution of the extremes, and we therefore
propose more complex models. We begin with an exploratory analysis
highlighting some of the peculiarities of the data, and then in Section
\ref{sec:spat_max} present the max-stable processes of \citet
{schlather02} and \citet{smith91}, which are extended in Section
\ref
{sec:maxstab_snow} to our extreme snow depth data. As full likelihood
inference is impossible for such models, in Section \ref
{sec:estimation} we discuss how composite likelihood inference may be
used for model estimation and comparison. The results of the data
analysis are presented in Section \ref{sec:appli} and a concluding
discussion is given in Section \ref{sec:discussion}.

%f1 ###
\begin{figure}[b]

\includegraphics{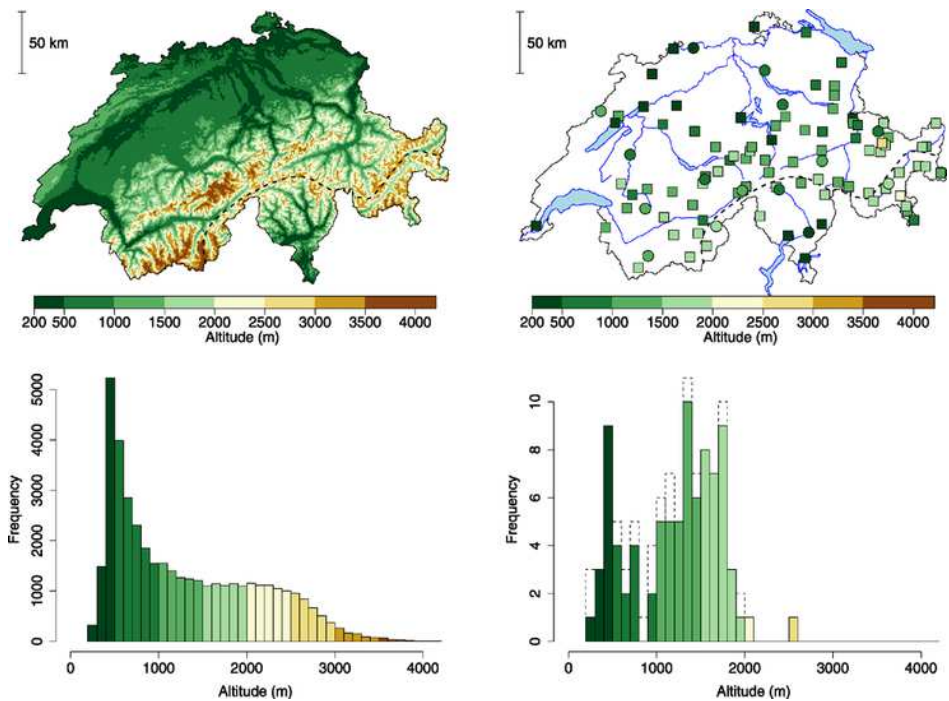}

\caption{Topography and locations of stations for which daily snow
depth data are available. First row: Topographical map of Switzerland
(left) and station locations (right). Second row: Histogram of
elevation of Switzerland at a $1$ km grid (left) and of the stations (right).
Color indicates altitude in meters above mean sea level. Among the
$101$ stations, $15$ (denoted by circles in the map on the right and by
the dashed part of the right-hand histogram) are excluded from the
analysis for validation. Dashed lines in the maps delimit the northern
and southern slopes of the Alps.}\label{fig:map_swiss}
\end{figure}

%s2 ###
\section{Preliminaries}

%s2.1 ###
\subsection{Data}\label{sec:data}

We consider annual maximum snow depth from the $101$ stations whose
locations are shown in Figure \ref{fig:map_swiss}. The stations belong
to two networks run by the WSL Institute for Snow and Avalanche\vadjust{\eject}
Research (SLF) and the Swiss Federal Office for Meteorology and
Climatology (MeteoSwiss). Annual maxima are extracted from daily snow
depth measurements, which are read off a measuring stake at around 7.30
AM daily from November $1$st to April $30$th, for the 43 winters
1965--1966 to 2007--2008; we use the term ``winter $1966$'' for the months
November $1965$ to April $1966$, and so forth. Examples of such time
series can be found in the Supplementary Materials, \citet
{BlanchetDavison2011Supp}. As Figure \ref{fig:map_swiss} shows, the
stations are denser in the Alpine part of the country, which has high
tourist infrastructure and increased population density and traffic
during the winter months. Their elevations range from $250$ m to $2500$
m above mean sea level, with only two stations above $2000$ m. In order
to validate our final model, we used $86$ stations to choose and fit
the model and retained 15 stations for model validation.

%s2.2 ###
\subsection{Marginal analysis and transformation}\label{sec:marginal}

Let $Z(x)$ denote the annual maximum snow depth at station $x$ of the
set $\mathcal X$, which here denotes Switzerland. %There are obviously
%very strong spatial correlations of annual maximum snow depths; $Z(x)$
%should therefore be treated as a spatial process.
Data are only available at the stations $x\in\mathcal D \subset
\mathcal X$, so modeling $Z(x)$ involves inference for the joint
distribution of $\{Z(x), x \in\mathcal X\}$ based on observations from
$\mathcal D$, and extrapolation to the whole of $\mathcal X$. In
particular, as the station elevations lie mainly below $2000$ m, any
results must be extrapolated to elevations higher than $2000$ m.

Daily snow depths at a given location $x$ are obviously temporally
dependent. However, time series analysis suggests that, for every
location $x\in\mathcal D$ and every winter, daily snow depths show
only short-range dependence. Hence, distant maxima of daily snow depths
seem to be near-independent and, therefore, the $D(u_n)$ condition for
independence of extremes that are well separated in time [\citet
{leadbetter83}, Section 3.2] should be satisfied. Extreme value theory
is then expected to apply to annual maximum snow depth: $Z(x)$ at a
location $x$ may be expected to follow a generalized extreme-value
(GEV) distribution [\citet{coles01}]
%
%e1 ###
\begin{equation}\label{eq:transf_frechet}
G(z)=\exp\biggl[- \biggl\{1+\xi(x)\frac{z-\mu(x)}{\sigma(x)} \biggr\}_{+}^{-1/\xi
(x)} \biggr],
\end{equation}
where $u_+=\max(u,0)$ and $\mu(x)$, $\sigma(x)>0$ and $\xi(x)$ are,
respectively, location, scale and shape parameters.

Characterizing the probability distribution of $Z(x)$ for all $x \in
\mathcal X$ is equivalent to characterizing the probability
distribution of $f\{Z(x)\}$ for any bijective function $f$, which may
be easier for a well-chosen $f$. A first step in our analysis is to
transform the data at the stations to the unit Fr\'{e}chet scale.
Whatever the values of the GEV parameters $\mu(x)$, $\sigma(x)$ and
$\xi
(x)$, taking $f(z)=-1/\log G(z)$ transforms $\{Z(x), x \in\mathcal X\}
$ into a spatial process $\{Z^*(x), x \in\mathcal X\}$ having unit Fr\'
{e}chet marginal distributions, $G^*(z)=\exp(-1/z)$. As it is easier to
deal with $Z^*$ in general discussion, we will assume below that the
time series at each station has been\vadjust{\eject} transformed in this way. To do so,
one might model the GEV parameters $\mu(x)$, $\sigma(x)$ and~$\xi(x)$
as smooth functions of covariates indexed by $x$, such as longitude,
latitude and elevation [\citet{padoan10}]. However, due to the very
rough topography of Switzerland and the influence of meteorological
variables such as wind and temperature, snow depth exhibits strong
local variation and additional covariates are necessary. A systematic
discussion of such covariates and associated smoothing is given by
\citet
{blanchet10}. The focus in the present paper is spatial dependence, so
rather than adopt their approach, here we simply use GEV fits for the
individual stations to transform $Z(x)$ at station $x\in\mathcal D$
into $Z^*(x)$. Diagnostic tools such as QQ-plots showed a good fit even
at low altitudes.

%s2.3 ###
\subsection{Spatial dependence and regional patterns}\label{sec:spatial_dpce}

\begin{figure}

\includegraphics{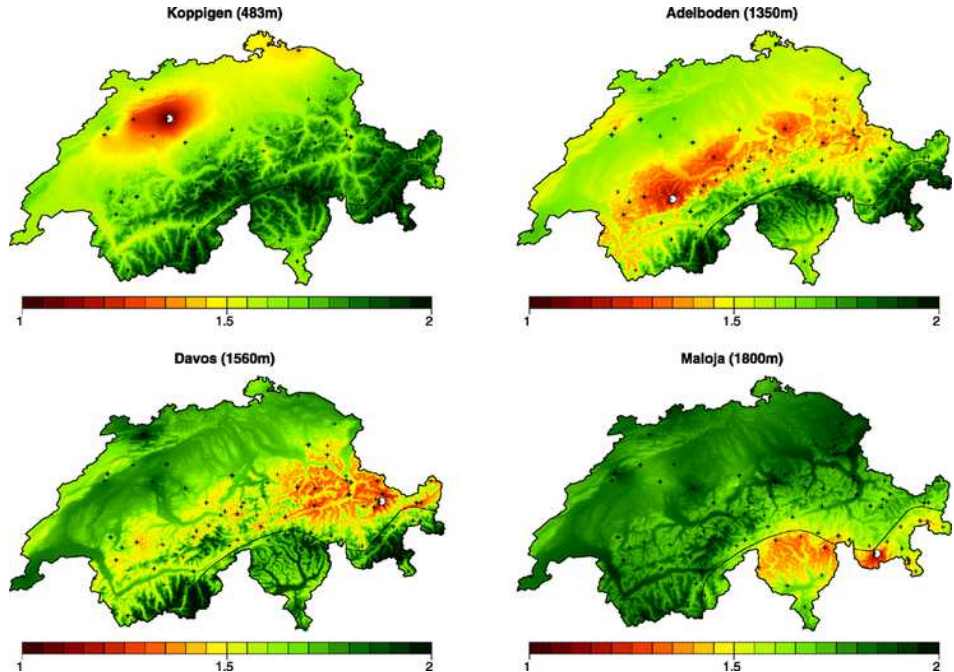}

\caption{Extremal coefficient computed relative to Koppigen, Adelboden,
Davos and Maloja (white points), estimated by the Cooley, Naveau and Poncet (\protect\citeyear{cooley06b})
madogram estimator, and then kriged to the whole of Switzerland using a
linear trend on absolute altitude difference.}\label{fig:krigmap_extcoeff}
\end{figure}

A simple measure of the dependence of spatial maxima at two stations
$x,x'\in\mathcal X$ is the extremal coefficient $\theta_{xx'}$. If
$Z^*(x)$ is the limiting process of maxima with unit Fr\'{e}chet
margins, then [\citet{coles01}, Chapter 5]
%
%e2 ###
\begin{equation}\label{eq:extrcoeff}
\pr\{Z^*(x)\leq z, Z^*(x')\leq z \}= \exp(-\theta_{xx'}/z),\qquad  z>0.
\end{equation}
One interpretation of $\theta_{xx'}$ appears on noting that
\[
\pr\{Z^*(x')>z| Z^*(x)>z \} \to2-\theta_{xx'},\qquad  z\to\infty.
\]
If $\theta_{xx'}=1$, then the maxima at the two locations are perfectly
dependent, whereas if $\theta_{xx'}=2$, they are asymptotically
independent as $z\to\infty$, so very rare events appear independently
at the two locations. Although they do not fully characterize
dependence, such coefficients are useful summaries of the
multidimensional extremal distribution. In particular, it may be
informative to compute all extremal coefficients $\{\theta_{xx'}, x'
\in\mathcal X\}$ for a given station $x$ to see how extremal
dependence varies. Figure \ref{fig:krigmap_extcoeff} depicts such maps
for the snow depth data, for four different reference stations $x$.
Extremal coefficients $\{\theta_{xx'}, x' \in\mathcal D\}$ were
estimated by the madogram-based estimator of \citet{cooley06b},
and then
kriged to the entire area using a linear trend on absolute altitude
difference between $x$ and~$x'$. Similar maps have been proposed for
gridded data by \citet{Coelhoetal2008}.

%f2 ###

Much information can be gleaned from Figure \ref{fig:krigmap_extcoeff}.
A strong elevation effect is clearly visible. The map for Adelboden
also suggests a directional effect: for this mid-altitude station in
the Alps, there is more dependence with other middle-altitude stations
in a roughly north-easterly direction. Another striking feature visible
in the two lower maps is near-independence between the northern and
southern slopes of the Alps. Further such maps suggest the presence of
the two weakly dependent regions separated by the black dotted line in
Figure \ref{fig:map_swiss}. A similar north/south separation was seen
in \citet{blanchet09}, for good reason: extreme snowfall events
occurring in these two regions typically do not stem from the same
precipitation systems. Whereas extreme snowfall events on the northern
slope of the Alps usually\vadjust{\eject} arise from northerly or westerly airflows
[\citet{schuepp78}], those in the southern slope usually come
from the south or south-west. These are less frequent, but when they occur they
can be very severe, due to the proximity of the Mediterranean Sea. As
snow cover results from the accumulation of many snowfall events during
the winter, one can expect annual maximum snow depths on the northern
and southern slopes of the Alps to be somewhat disconnected. The winter
of $1981$ illustrates this: little snow fell on the southern slope of
the Alps, while the northern slope received large amounts. Figure \ref
{fig:krigmap_extcoeff} nevertheless suggests that these two regions are
asymptotically weakly dependent, since $\theta_{xx'}$ is generally
larger than $1.7$, but not necessarily asymptotically independent. Even
between well-separated stations, $\theta_{xx'}$ is rarely very close to
2, perhaps owing to the rather small area under study, in which the
largest distance between stations is around $350$ km.

%s3 ###
\section{Spatial maxima}\label{sec:spat_max}

%s3.1 ###
\subsection{Max-stable processes}
\label{sec:maxstab}

The spatial dependence highlighted in Section~\ref{sec:spatial_dpce}
suggests that we model $Z^*(x)$ as a spatial process of extremes.
%Max-stable representations characterize such processes.
A {max-stable} process with unit Fr\'{e}chet margins is a stochastic
process $\{Z^*(x), x \in\mathcal X\}$ with the property that, if
$Z^*_{(1)}(x), \ldots, Z^*_{(n)}(x)$ are $n$ independent copies of the
process, then [\citet{dehaan84}]
\[
\Bigl\{\max_{i=1,\ldots,n} Z^*_{(i)}(x), x \in\mathcal X \Bigr\} \mbox{ has the
same distribution as } \{n Z^*(x), x \in\mathcal X \}.
\]
A consequence of this definition is that all finite-dimensional
marginal distributions are max-stable: if $\{x_1, \ldots, x_D\}$ is a
finite subset of $\mathcal X$, then for all $n\in\mathbb{N}$,
\begin{eqnarray*}
&&\pr\{Z^*(x_1) \leq n z_1, \ldots, Z^*(x_D) \leq n z_D\}^n\\
&&\qquad =\pr\{Z^*(x_1)
\leq z_1, \ldots, Z^*(x_D) \leq z_D\},\qquad  z_1, \ldots, z_D>0.
\end{eqnarray*}
Such processes have several representations, two of which we now sketch.
%There are several representations of such processes. We present two of
%them \citep{dehaan84}, \citet{schlather02})
%in the next sections.

%s3.2 ###
\subsection{Smith's storm model}\label{sec:smith_model}

A general method of constructing max-stable processes is due to \citet
{dehaan84}. Let $\{(\eta_i,s_i), i \in\mathbb{N}\}$ denote the points
of a Poisson process on $(0,\infty) \times\mathcal S$ with intensity
% measure
$\eta^{-2} d\eta\times\nu(ds)$, where $\mathcal S$ is an arbitrary
measurable set and $\nu$ is a positive measure on $\mathcal S$. Let $\{
f(s,x), s \in\mathcal S, x \in\mathcal X\}$ denote a nonnegative
function for which, for all $x\in\mathcal X$,
\[
\int_{s \in\mathcal S} f(s,x) \nu(ds)=1.
\]
Then the random process
%
%e3 ###
\begin{equation}\label{eq:constr_dehaan}
Z^*= \Bigl\{\max_{i \in\mathbb{N}} \{ \eta_i f(s_i,x)\}, x \in\mathcal X
\Bigr\}
\end{equation}
is max-stable with unit Fr\'{e}chet margins. \citet{smith91}
gives a
rainfall-storms interpretation of this construction. He suggests
regarding $\mathcal S$ as a space of storm centers, of $f(s,\cdot)$ as
the shape of a storm centered at $s$, and of $\eta$ as a~storm
magnitude. Then $\eta f(s,x)$ represents the amount of rainfall
received at location $x$ for a storm of magnitude $\eta$ centered at
$s$ and $Z^*(x)$ in (\ref{eq:constr_dehaan}) is the maximum rainfall
received at $x$ over an infinite number of independent storms.

Additional assumptions are needed to get useful models from (\ref
{eq:constr_dehaan}). \citet{smith91} proposes taking $\mathcal
S=\mathcal X=\mathbb{R}^D$, letting $\nu$ be the Lebesgue measure and
$f(s,\cdot)$ be a multivariate normal density with mean $s$ and
covariance~ma\-trix~$\Sigma$, that is,
\[
f(s,x)=(2 \pi)^{-D/2} |\Sigma|^{-1/2} \exp\bigl\{-\tfrac{1}{2} (x-s)^T
\Sigma^{-1} (x-s) \bigr\},\qquad  x, s \in\mathbb{R}^D.
\]
The resulting bivariate distribution of $Z^*$ defined by (\ref
{eq:constr_dehaan}) at two stations $x_1$ and $x_2$ is then
%
%e4 ###
\begin{eqnarray}\label{eq:biv_smith}
&&\pr\{Z^*(x_1) \leq z_1, Z^*(x_2) \leq z_2 \}\nonumber\\[-8pt]\\[-8pt]
&&\qquad =\exp\biggl\{
-\frac{1}{z_1} \Phi\biggl(\frac{a}{2}+\frac{1}{a}\log\frac{z_2}{z_1} \biggr)
-\frac{1}{z_2} \Phi\biggl(\frac{a}{2}+\frac{1}{a}\log\frac{z_1}{z_2} \biggr)
\biggr\},\nonumber
\end{eqnarray}
where $\Phi$ is the standard normal distribution function and $a$ is
the Mahalano\-bis distance given by
%
%e5 ###
\begin{equation}\label{eq:mahaldist}
a^2={(x_1-x_2)^T \Sigma^{-1} (x_1-x_2)}.
\end{equation}
Below we will call this model the Smith process.

%f3 ###
\begin{figure}[t]

\includegraphics{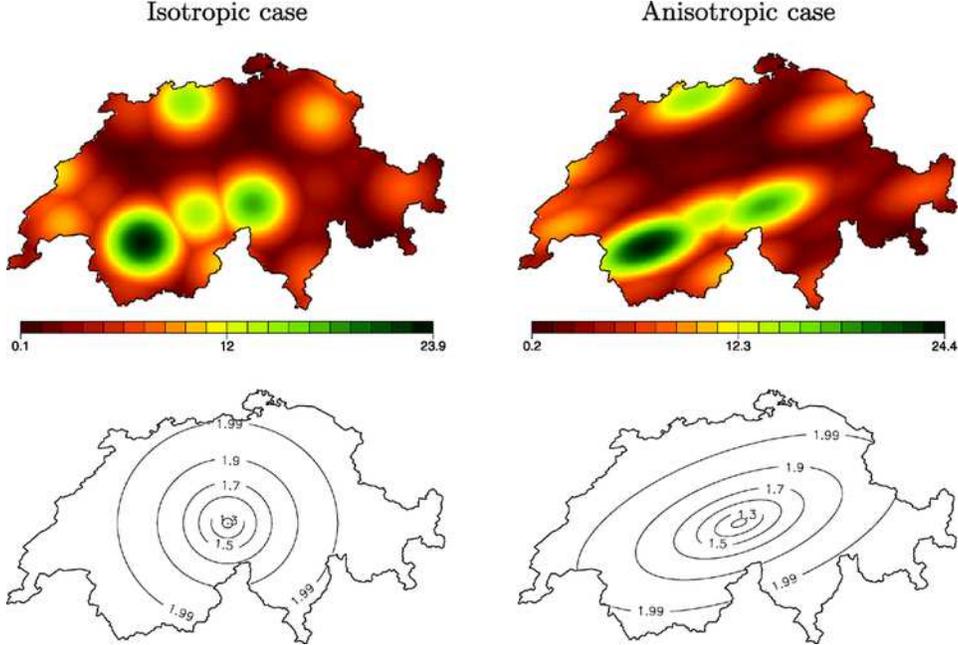}%
\vspace*{-2pt}
\caption{Smith's process in two dimensions with two different matrices
$\Sigma=(\tau_{dd'})_{d,d'\in\{1,2\}}$. Upper left image: a simulated
field with $\tau_{11}=\tau_{22}=17^2$ and $\tau_{12}=0$ (isotropic
case). Upper right image: a simulated field with $\tau_{11}=25^2$,
$\tau
_{22}=15^2$ and $\tau_{12}=14^2$ (anisotropic case). Lower images:
corresponding pairwise extremal coefficient.}
\label{fig:simu_smith}
\vspace*{-2pt}
\end{figure}

Two simulated Smith processes with different matrices $\Sigma$ are
shown in the top row of Figure \ref{fig:simu_smith}. The anisotropic
case arises when $\Sigma$ is not spherical, that is,\ not of the form
$\Sigma=\tau^2\mathbb{I}_D$, where $\tau^2>0$ and $\mathbb{I}_D$
is the
identity matrix of side $D$. The resulting geometric anisotropy [e.g.,
\citet{journel78}] can easily be seen by computing pairwise extremal
coefficients. Taking $z_1=z_2=z$ in (\ref{eq:biv_smith}) gives,
according to (\ref{eq:extrcoeff}),
%
%e6 ###
\begin{equation}\label{eq:extcoeff_smith}
\theta_{x_1x_2}= 2\Phi(a/2).
\end{equation}
The Mahalanobis distance $a$ appearing in (\ref{eq:extcoeff_smith})
gives different weights to the different components of the vector
$(x_1-x_2)$. The limiting cases $a \rightarrow0^+$ and $a \rightarrow
+\infty$ correspond, respectively, to perfect dependence, $\theta
_{x_1x_2}=1$, and independence, $\theta_{x_1x_2}=2$. For a given
station $x_1$, surfaces \mbox{$\{x_2 \in\mathcal X, \theta_{x_1 x_2}=c\}$}
are, according to (\ref{eq:extcoeff_smith}), such that (\ref
{eq:mahaldist}) is constant.
If $\Sigma$ is spherical, then such surfaces are circles in two
dimensions and spheres in three dimensions. Otherwise, they are
ellipses and ellipsoids, respectively.\vadjust{\eject}

%s3.3 ###
\subsection{Schlather's storm model} \label{sec:schlather_model}

A second method of construction of max-stable processes was proposed by
\citet{schlather02}. Let $\{\eta_i, i \in\mathbb{N}\}$ denote the
points of a Poisson process on $\mathbb{R}_+ $ with intensity
% measure
$\eta^{-2} d\eta$. Let $\{W(x), x \in\mathcal X\}$ be a stationary
nonnegative process that satisfies $\mathbb{E}[W(x)]=1$ for all $x
\in
\mathcal X$, and let $W_i$, $i \in\mathbb{N}$, be independent copies
of this process. Then [\citet{schlather02}] the random process
%
%e7 ###
\begin{equation}\label{eq:constr_schlather}
Z^*= \Bigl\{ \max_{i \in\mathbb{N}} \eta_i W_i(x), x \in\mathcal X \Bigr\}
\end{equation}
is max-stable with unit Fr\'{e}chet margins. When $W_i(x)=f(x-s_i)$,
where $f$~is a density function on $\mathcal X$ and the $s_i$ are the
points of a Poisson process with unit rate on a measurable set
$\mathcal S$, then \eqref{eq:constr_schlather} is equivalent to the
storm model of Section~\ref{sec:smith_model}. Smith's model (\ref
{eq:biv_smith}) corresponds to taking $f$ to be a multivariate normal
density, extended by \citet{dehaan06} to Student $t$ and Laplace
densities. Like Smith's model, the model (\ref{eq:constr_schlather})
has a simple interpretation: the $\eta W$ are spatial events all having
the same stochastic dependence structure but differing in their
magnitudes $\eta$. An appealing difference between this and the Smith
model is that the shapes of the events may vary if the process $W$
permits this.

Additional assumptions are again needed to get useful models from (\ref
{eq:constr_schlather}). \citet{schlather02} proposes taking $W$
to be
the positive part of a stationary Gaussian process with correlation
function $\rho$, scaled so that $\mathbb{E}[\max\{0,\break W(x)\}]=1$ for all
$x \in\mathcal X$. He shows that the corresponding bivariate
distribution of $Z^*$ at two stations $x_1$ and $x_2$ is
%
%e8 ###
\begin{eqnarray}\label{eq:biv_schlather}
&&\pr\{Z^*(x_1) \leq z_1, Z^*(x_2) \leq z_2\}\nonumber\\[-8pt]\\[-8pt]
&&\qquad =\exp\biggl\{-\frac{1}{2} \biggl(\frac{1}{z_1}+\frac{1}{z_2} \biggr)
\biggl(1+\sqrt{1-2 (\rho(h)+1) \frac{z_1 z_2}{(z_1+z_2)^2}} \biggr)\biggr\},\nonumber
\end{eqnarray}
where $h \in\mathbb{R}_{+}$ is the Euclidean distance $\|x_2-x_1\|$
between the two stations. Below we call this max-stable model
Schlather's process.

%f4 ###
\begin{figure}[t]

\includegraphics{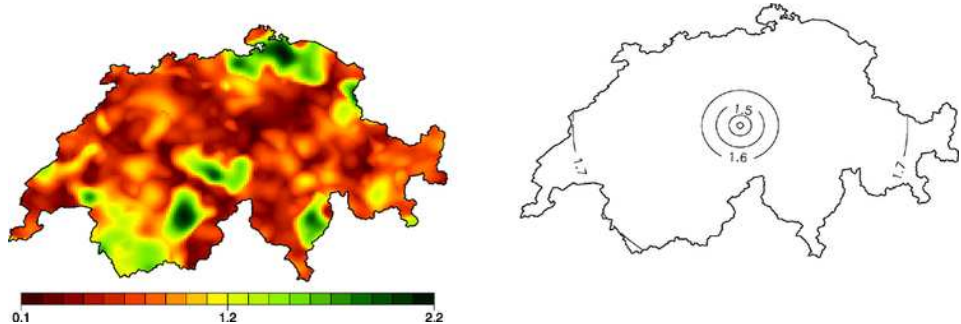}%
\vspace*{-2pt}
\caption{Schlather's process in two dimensions for a Cauchy covariance
function $\rho(h)=(1+h^2/19^2)^{-1}$. Left image: one simulated field.
Right image: corresponding pairwise extremal coefficient.}
\label{fig:simu_schlather}
\vspace*{-2pt}
\end{figure}

A simulation from an isotropic version of this model with $\mathcal X$
corresponding to Switzerland is shown in Figure \ref
{fig:simu_schlather}. The isotropy can be easily seen by computing
pairwise extremal coefficients. Taking $z_1=z_2=z$ in (\ref
{eq:biv_schlather}) gives, according to (\ref{eq:extrcoeff}),
%
%e9 ###
\begin{equation}\label{eq:extcoeff_schlather}
\theta_{x_1x_2}=1+ \biggl\{\frac{1-\rho(\|x_1-x_2\|)}{2} \biggr\}^{1/2}.
\end{equation}
Here the extremal coefficients involve the Euclidean distance between
the two locations. For a given station $x_1$, surfaces with the same
extremal coefficents $c \in[1,2]$, that is,\ surfaces $\{x_2 \in
\mathcal X, \theta_{x_1 x_2}=c\}$, are, according to (\ref
{eq:extcoeff_schlather}), such that
$\|x_1-x_2\| = c'$.
Such surfaces are circles in two dimensions and spheres in three
dimensions. The limiting case $\|x_1-x_2\| \rightarrow0^+$ corresponds
to perfect dependence, $\theta_{x_1x_2}=1$. If, like most
geostatistical correlation functions, the underlying Gaussian process
has $\rho(h)\rightarrow0$ when $h \rightarrow+\infty$,\vadjust{\eject} then the
limiting case $\|x_1-x_2\| \rightarrow+\infty$ corresponds to $\theta
_{x_1x_2}=1+2^{-1/2} \approx1.707$, and so independent extremes do not
arise even at very large distances. Moreover, as an isotropic
correlation function can give correlations no smaller than $-0.403$ in
$\mathbb{R}^2$ and $-0.218$ in $\mathbb{R}^3$ [\citet
{matern86}, page
16], under Schlather's model we have $\theta_{x_1 x_2} \leq1.838$ for
any $x_1$, $x_2$ in $\mathbb{R}^2$ and $\theta_{x_1 x_2} \leq1.780$
for any $x_1$, $x_2$ in $\mathbb{R}^3$. Thus, it is impossible to
produce independent extremes using such a process, no matter how
distant the stations. \citet{davison10} have proposed extensions to
allow independence in (\ref{eq:biv_schlather}), and \citet{kabluchko09}
have extended both Smith's and Schlather's representations.

%s4 ###
\section{Max-stable process for extreme snow depth}\label{sec:maxstab_snow}

%s4.1 ###
\subsection{General}

As pointed out in Section \ref{sec:spatial_dpce}, snow depth data show
two key characteristics that should be explicitly modeled in the
max-stable process. First, dependence is anisotropic, due to the strong
elevation effect and the presence of a main direction of dependence.
Second, Switzerland seems to be divided into two weakly dependent
climatic regions: the northern slope of the Alps together with the
Plateau, which is the low altitude region north of the Swiss Alps; and
the southern slope of the Alps. In this section we propose to extend
the Smith and Schlather models of Sections \ref{sec:smith_model}--\ref
{sec:schlather_model} to account for these features. Other
representations described in \citet{kabluchko09}, in \citet
{davison10} or
in \citet{DavisonPadoanRibatet2010} are not considered in this paper.

%s4.2 ###
\subsection{Modeling anisotropy} \label{sec:anisotrop}

Smith's model can directly model anisotropy using a nonspherical
$\Sigma
$ matrix in (\ref{eq:biv_smith}). The simple version of Schlather's
model is isotropic, but it can easily account for anisotropy by
considering a~transformed space $\mathcal{\tilde X}$ instead of
$\mathcal X$.

Anisotropy of Smith's model arises from the fact that the distance used
in the extremal coefficient (\ref{eq:extcoeff_smith}) is Mahalanobis
distance (\ref{eq:mahaldist}) rather than Euclidean distance. Using the
eigendecomposition $\Sigma=U \Lambda U^T$, where $U$ is a~rotation
matrix and $\Lambda$ a diagonal matrix of positive eigenvalues, we may write
%
%e10 ###
\begin{equation}\label{eq:isigma}
\Sigma^{-1}=U^T \Lambda^{-1} U = (\Lambda^{-1/2} U)^T (\Lambda^{-1/2} U),
\end{equation}
where $\Lambda^{-1/2}$ denotes the diagonal matrix composed of the
reciprocal square roots of the diagonal elements of $\Lambda$. If
$\lambda_{1}$ denotes the first element of $\Lambda$, then (\ref
{eq:isigma}) can be written as
$\Sigma^{-1}=\lambda_{1}^{-1} V^T V$, where $V=\lambda
_{1}^{1/2}\Lambda
^{-1/2} U$. The squared Mahalanobis distance (\ref{eq:mahaldist}) is
\[
a^2=\frac{1}{\lambda_{1}} (x_1-x_2)^T V^T V (x_1-x_2) = \frac
{1}{\lambda_{1}} [V (x_1-x_2) ]^T [V (x_1-x_2) ],
\]
which is exactly that between $\tilde x_1=V x_1$ and $\tilde x_2=V x_2$
in the isotropic case, that is,\ when using a $D$-dimensional spherical
covariance matrix $\lambda_{1} \mathbb{I}_D$ in (\ref{eq:biv_smith}).
Thus, the anisotropic Smith model on $\mathcal X$ is just the isotropic
Smith model on the transformed space $\mathcal{\tilde X}= V \mathcal X$.

%f5 ###
\begin{figure}[b]

\includegraphics{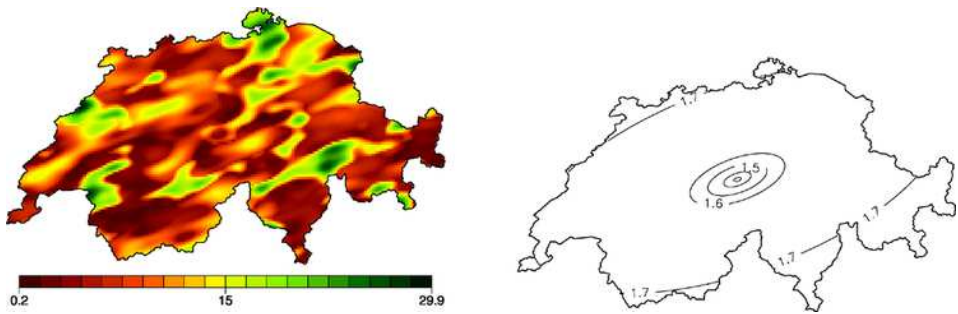}

\caption{Anisotropic Schlather model resulting from climate space
transformation. Left image: a simulated field. Right image:
corresponding extremal coefficients.}\label{fig:geogr_clim_space}
\end{figure}

Similar ideas can be used with Schlather's model, by applying it on
$\mathcal{\tilde X}= V \mathcal X$, where in three dimensions we may take
%
%e11 ###
\begin{equation}\label{eq:V_3D}
V= \pmatrix{
\cos\alpha& -\sin\alpha& 0\cr
c_2 \sin\alpha& c_2 \cos\alpha& 0\cr
0& 0& c_3
} ,\qquad  c_2, c_3 \in\mathbb{R}_+^*,
\end{equation}
as for Smith's model.
In the rest of the paper we will use the term climate space for~the
transformed space $\mathcal{\tilde X}= V \mathcal X$ in which isotropy
is achieved. Figure \ref{fig:geogr_clim_space} illustrates the climate
space transformation, allowing an anisotropic Schlather model, with the
same $V$ matrix as that corresponding to the anisotropic case of
Figure~\ref{fig:simu_smith}. Compared to Figure \ref{fig:simu_schlather},
constant extremal coefficients correspond to ellipses, allowing us to
model directional effects.

Geometric anisotropy as induced by the $V$ matrix is a special case
of~ran\-ge anisotropy [\citet{zimmerman1993}]. In the nonextremal framework,
this idea has been extended to nongeometric range anisotropic models,
in which nested covariances are used with different range parameters in
different directions, but in general this does not define a valid
covariance function. \citet{eckertgelfand2003} introduced product
geometric anisotropy, under which covariance functions are products of
geometric anisotropic covariances. Space transformation has also been
used by \citet{guttorp92} to model nonstationary spatial covariance
structures, allowing more complex transformations than the affine
transformation considered here. In addition to these global methods,
local methods for modeling anisotropy and more general forms of
nonstationarity also exist. These can be divided in three main families
[\citet{SchabenbergerGotway2005}]. The moving window approach of
\citet
{haas1990} estimates a covariance function locally within a
neighborhood. The convolution method of \citet{higdon1998} allows the
construction of weakly nonstationary processes by convolving a
zero-mean white noise process with a kernel function whose parameters
can depend on location. The method of weighted stationary processes
[\citet{fuentes2001}]\vadjust{\goodbreak} allows one to write the nonstationary covariance
function as a weighted mixture of isotropic covariances, where the
weights depend on the location. \citet{FuentesHenryReich2010} use a
Dirichlet process mixture as the basis for a~flexible copula approach
to space-time modeling of extreme temperatures, but it does not
correspond to a max-stable process model, and the relatively long-range
dependence of temperatures can be modeled more simply than can
precipitation phenomena such as rain- and snowfall. It would be very
valuable to apply these ideas in the max-stable context, but the
unavailability of a likelihood function seems to be a major obstacle.

The idea of space transformation was used by \citet{cooley07} in
modeling US precipitation. Instead of using the three-dimensional
geographical coordinates (longitude, latitude, elevation) for locating
stations, the authors work in a ``climate space,'' namely, the
two-dimensional space given by elevation and mean precipitation for the
months April to October. Unlike in \citet{cooley07}, our transformation
is affine, giving more weight to elevation through $c_3$, and defining
a main direction of dependence along the $\alpha$-axis. A
higher-dimensional space could of course be used for $\mathcal X$. In
particular, one could use the four-dimensional space of (longitude,
latitude, elevation, mean snow depth), thus blending the \citet
{cooley07} approach with ours; see Section \ref{sec:appli}.

%s4.3 ###
\subsection{Modeling climate regions}\label{sec:model_region}

\begin{figure}

\includegraphics{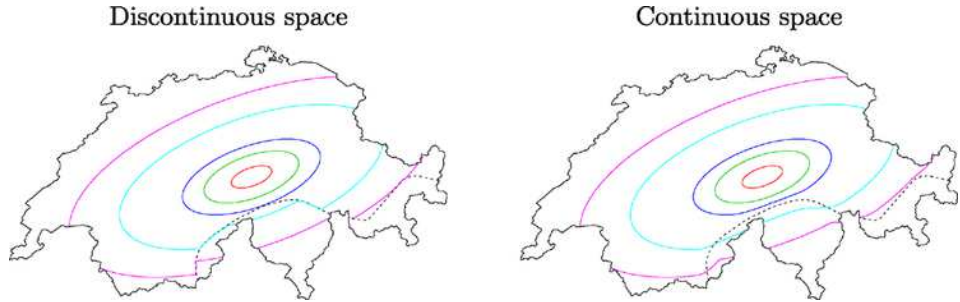}

\caption{Example of extremal coefficients with weakly dependent regions
in discontinuous space (left image) and continuous space (right image).
The images are the same, except in a 10 km wide band around the
north/south border (dashed line).}\label{fig:extrcoeff_climreg}
\end{figure}

Different approaches to accounting for the impact of the climate
regions on the extremes are possible:
\begin{enumerate}
\item\textit{The climate regions are independent.} This is equivalent
to saying that two max-stable processes govern the two regions
independently. In terms of spatial dependence, extremal coefficient
maps will be of the form of Figure~\ref{fig:simu_smith} or \ref
{fig:geogr_clim_space} but replacing the Swiss border by the border of
the northern region alone for the pairwise dependence with a station
located in the north, and similarly for the southern region.

\item\textit{The climate regions are weakly dependent.} Since
dependence between pair of stations decreases when distance increases,
one way to model weakly dependent regions is to increase the distance
between them. This can be done by adding to $\mathcal X$ a coordinate
equal to $0$ in the northern region, and to $1$ in the southern region.
If the other coordinates are (longitude, latitude, elevation), then the
$V$ matrix of the climate space transformation (\ref{eq:V_3D}) can be
written in the most general case as a $4 \times4$ matrix with one
column comprising $0$ apart from one element. Nevertheless, for
computational reasons it may be better to consider the rotation matrix
$U$ of Section \ref{sec:anisotrop} as being a rotation matrix in the
(longitude, latitude) plane and thus to set\vspace*{3pt}
%
%e12 ###
\begin{equation}\label{eq:V_4D}
V= \pmatrix{
\cos\alpha& -\sin\alpha& 0 & 0\cr
c_2 \sin\alpha& c_2 \cos\alpha& 0 & 0\cr
0& 0& c_3 & 0\cr
0 & 0 & 0 & c_4
};\vspace*{3pt}
\end{equation}
we shall do this in Section \ref{sec:appli}. In the four-dimensional
climate space $\mathcal{\tilde X}=V \mathcal X$, the squared distance
between two stations $x_1$ and $x_2$ is $\{V(x_1-x_2)\}^T \{V(x_1-x_2)\}
$. But the fourth coordinate of $x_1-x_2$ is $0$ if the two stations
belong to the same region and $\pm1$ otherwise. The squared distance
will then equal that in the (longitude, latitude, elevation) climate
space if the two stations are in the same region, and be increased by
$c_4^2$ otherwise. We thus increase the distance between the climate
regions, and therefore decrease the dependence between them, without
increasing the distance between stations of the same region. To see how
the extremal coefficients behave, see the left-hand side of Figure \ref
{fig:extrcoeff_climreg}.

\item\textit{The climate regions are weakly dependent in continuous
space.} Since the additional coordinate introduced above jumps from $0$
to $1$ at the border between the regions, it induces a discontinuity of
the extremal coefficients which is visible in the left map of Figure
\ref{fig:extrcoeff_climreg}; see the cyan and magenta ellipses. This
seems unrealistic and something smoother is preferable. An easy way to
impose space continuity is to take the border to be a~band inside which
the fourth coordinate is linearly interpolated between $0$ and~$1$,
with value $0$ on the upper-border of the band and $1$ on the
lower-border. With this simple interpolation, there is no jump at the
border and curves of constant extremal coefficient are continuous, as
in the right-hand side of Figure \ref{fig:extrcoeff_climreg}. The width
of the band must be estimated from the data; we return to this in
Section \ref{sec:estimation}.
\end{enumerate}

%f6 ###

%s5 ###
\section{Model estimation and selection}\label{sec:estimation}

%s5.1 ###
\subsection{Pairwise likelihood}\label{sec:pairlik}

Statistical inference for parametric models is ideally performed using
the likelihood function. Let $\mathcal D=\{x_1, \ldots, x_D\}\subset
\mathcal X$ denote the $86$ stations whose maxima are used for fitting
the models. Computation of the likelihood requires the joint density
function of $\{Z^*(x_1), \ldots, Z^*(x_D)\}$, but in the framework of
max-stable processes, this is infeasible because only the bivariate
marginal distributions are available. \citet{padoan10} proposed
replacing the full likelihood by a pairwise likelihood function [\citet
{cox04}; \citet{varin08}]. This idea is also used by \citet{davison10},
\citet
{DavisonPadoanRibatet2010} and by \citet{smith09}, the latter in a
Bayesian framework.

Let $z_{ik}$ denote the $k$th observed maximum for the $i$th station,
transformed so that time-series $(z_{i1}, \ldots, z_{iK})$ at each
station have unit Fr\'{e}chet distributions; here $k \in\{1, \ldots,
K\}$, with $K=43$ years, and $i \in\{1,\ldots,D\}$, with $D=86$
stations. Let ${\beta}=(\beta_1, \ldots, \beta_R)$ denote the
parameters to be estimated. Then the pairwise marginal log-likelihood is
%
%e13 ###
\begin{equation}
\ell_{\mathrm{p}}({\beta})=\sum_{k=1}^K \sum_{i<j} \log f(z_{ik},
z_{jk};{\beta}),
\label{eq:pair_lik}
\end{equation}
where $f(\cdot,\cdot)$ is the bivariate density of the unit Fr\'{e}chet
max-stable process, that is,\ the derivative of equation (\ref
{eq:biv_smith}) for Smith's model or of (\ref{eq:biv_schlather}) for
Schlather's model, and the second summation is over all distinct pairs
of stations, $D(D-1)/2$ terms in all. Under suitable regularity
conditions, the maximum pairwise maximum likelihood estimator $\check
{\beta}$ has a limiting normal distribution as $K \rightarrow+\infty$,
with mean ${\beta}$ and covariance matrix of sandwich form estimable by
$H({\check{\beta}})^{-1} J({\check{\beta}})H({\check{\beta
}})^{-1}$, where
%
%e15 ###
%e14 ###
\begin{eqnarray}
H({\beta})&=&-\sum_{k=1}^K \sum_{i<j} \frac{ \partial^2
\log f(z_{ik}, z_{jk};{\beta})}{\partial{\beta}
\partial{\beta}^T}, \label{eq:J_matrix} \\
J({\beta})&=&\sum_{k=1}^K \sum_{i<j} \frac{ \partial
\log
f(z_{ik}, z_{jk};{\beta})}{\partial{\beta} }
\frac
{ \partial\log f(z_{ik}, z_{jk};{\beta})}{ \partial
{\beta}^T} \label{eq:K_matrix}
\end{eqnarray}
are the observed information matrix and the squared score statistic
corresponding to $\ell_{\mathrm{p}}$. The use of the pairwise likelihood
estimator for Smith's process was validated by \citet{padoan10}
in a
simulation study.

%s5.2 ###
\subsection{Estimation in practice} \label{sec:estim_practice}

Estimating the maximum pairwise likelihood estimator requires the
maximization of (\ref{eq:pair_lik}) with respect to the $R$~pa\-rameters.
We found that the \verb+R+ function \texttt{optim} gave quite poor
results for our application: the surface $\ell_{\mathrm{p}}$ can have many
local maxima, and \texttt{optim} and similar functions find it hard to
deal with them. After some experimentation, we therefore adopted a
profile likelihood method. Given a set of $(R-1)$ parameters
${\beta}_{-r}$, it is easy to find the value of $\beta_r$
that maximizes the single-variable function $\ell_{\mathrm{p}}(\cdot
,{\beta}_{-r})$. This suggests the following iterative algorithm:
\begin{enumerate}
\item Take initial parameters ${\beta}=(\beta_1,\ldots
,\beta_R)$.
\item For $r$ in $1, \ldots,R$:
\begin{enumerate}[(a)]
\item[(a)] find the value $\check{\beta}_r$ that maximizes the pairwise
likelihood with respect to the scalar $\beta_r$, holding the other
parameters, ${\beta}_{-r}$, fixed, that is,\
\[
\check{ \beta}_r=\arg\max_{\beta_r}\ell_{\mathrm{p}}(\beta_r,
{\beta}_{-r});
\]
\item[(b)] then update the $r$th component of ${\beta}$ to
$\check
{\beta}_r$.
\end{enumerate}
\item Go to step 2, stopping when no change to any $\beta_r$ can
increase the pairwise log-likelihood.
\end{enumerate}

%f7 ###
\begin{figure}[t]
\vspace*{-2pt}
\includegraphics{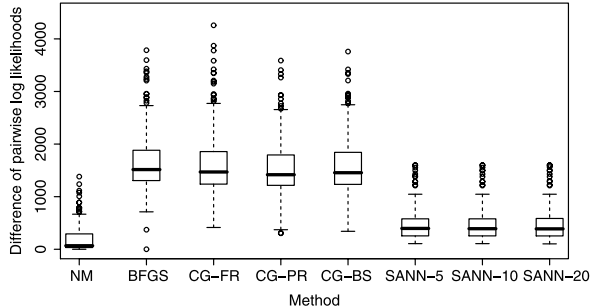}
\vspace*{-2pt}
\caption{Difference in pairwise log-likelihood at convergence between
the profiling algorithm and eight algorithms for simultaneous parameter
estimation, for 200 simulated data sets. The eight algorithms are:
Nelder--Mead, NM; the quasi-Newton method of Broyden, Fletcher,
Goldfarb and Shanno, BFGS; three conjugate gradient methods, CG-FR,
CG-PR and CG-BS; and a variant of simulated annealing using starting
temperatures of 5, 10 and 20, SANN-5, SANN-10 and SANN-20. The help for
the \texttt{R} function \texttt{optim} gives more details of these
algorithms.}\label{fig:compar_estim_meth}
\end{figure}

To assess the performance of this algorithm, we simulated 200 data
sets, each comprising $43$ independent copies of Schlather's max-stable
random field (\ref{eq:biv_schlather}) with Cauchy covariance function
$\rho$ in a three-dimensional climate space. Each of the copies is
observed at the same $D=100$ stations, so the number of observations is
very similar to those for the annual snow depth data; see Section \ref
{sec:data}. The climate transformation is defined through a $3 \times
3$ matrix $V$ as in (\ref{eq:V_3D}). The model has three parameters for
the $V$ matrix and two for the covariance function, which induces
middling dependence: about 25\% of the pairs of stations have extremal
coefficients $\theta\leq1.68$; recall from Section \ref
{sec:schlather_model} that for Schlather's model, $\theta\leq1.707$.
We started from the same initial point for each of the $200$ data sets
and optimized the log pairwise likelihood (\ref{eq:pair_lik}) with (i)
eight optimization procedures within\vadjust{\eject} the \texttt{R} function \texttt
{optim} with all parameters estimated jointly [\citet
{BlanchetDavison2011Supp}]; and (ii) the above profile likelihood
algorithm. Figure \ref{fig:compar_estim_meth} shows the differences
between the pairwise log-likelihoods for the methods at convergence,
for the 200 data sets. The profiling method never gives lower maximized
pairwise likelihoods than the other algorithms, and they are almost
always higher. Further simulations with small- and large-range
dependence gave similar results [\citet{BlanchetDavison2011Supp}]:
overall profiling is clearly better than the other algorithms. Those
that compare best with profiling, viz., Nelder--Mead and simulated
annealing, are designed for rather rough surfaces with many local
optima. These simulated data are relatively simple compared to the real
data, which are neither exactly unit Fr\'{e}chet after transformation
from (\ref{eq:transf_frechet}) nor follow a pure max-stable process.
Furthermore, the max-stable model used for the simulation is quite
simple, with only five parameters to be estimated, so the profiling
approach seems necessary for our, more complex, application.

%s5.3 ###
\subsection{Model selection}\label{sec:model_selec}

Model selection criteria play an important role in deciding which of
the fitted models should be preferred. As in \citet{padoan10}, we
propose to use the composite likelihood information criterion [\citet
{varin05}], which extends the $\TIC$ [\citet{takeuchi76}] to the
composite likelihood setting, and is defined as
\[
\CLIC=-2\ell_{\mathrm{p}}(\check{{\beta}}) + 2 \operatorname{tr} \{
H(\check
{{\beta}})^{-1} J(\check{{\beta}}) \},
\]
where $H$ and $J$ are, respectively, the observed information matrix and
the squared score statistic corresponding to $\ell_{\mathrm{p}}$, defined at
equations \eqref{eq:J_matrix} and~\eqref{eq:K_matrix}, and $\check
{{\beta}}$ is the maximum pairwise likelihood estimator.
Lower values of $\CLIC$ correspond to better quality models.

%s6 ###
\section{Application to snow depth in Switzerland} \label{sec:appli}

%s6.1 ###
\subsection{Fitted models} \label{sec:fitted_models}

We fitted the different models described in Section \ref
{sec:maxstab_snow} to our snow depth data, using both Smith and
Schlather max-stable structures for the extremes. For Schlather's
model, different choices of Gaussian covariance function $\rho$ lead to
different distributions (\ref{eq:biv_schlather}). We used nine such
functions, namely, the spherical, circular, cubic, Gneiting,
exponential, Mat\'{e}rn, Gaussian, powered-exponential and Cauchy
covariance functions [\citet
{BanerjeeCarlinGelfand2003}; \citet{SchabenbergerGotway2005}]. Each has either
one or two parameters and the first four have an upper bound. They all
are such that $\rho(h)\rightarrow1$ when $h \rightarrow0^+$ and
$\rho
(h)\rightarrow0$ when $h \rightarrow+\infty$. As mentioned in Section
\ref{sec:schlather_model}, this constrains the extremal coefficient for
Schlather's model to correspond to dependent data. Nevertheless, we
will see that such an assumption seems justified in our case.

The coordinates $x$ we considered are geographical coordinates
(longitude, latitude, elevation), region number (see Section \ref
{sec:model_region}) and mean snow depth during the winters 1966--2008.
Mean precipitation was considered as a possible climate coordinate in
\citet{cooley07}'s study of extreme precipitation. The idea of using
mean snow depth is that stations with similar snow depth are probably
influenced by the same weather patterns and should therefore be closer
in the climate space than are stations with different snow cover. Other
climate variables that could be considered are temperature, wind
direction and wind speed, which are also measured at the stations, but
these values are of relatively poor quality with many missing values, so
we decided not to use them.

In addition to the models illustrated in Section \ref
{sec:maxstab_snow}, we allowed the possibility of having different
climate spaces in northern and southern regions, that is,\ to have
different climate space transformation matrices $V$. In three
dimensions, for example, two $V$ matrices as in (\ref{eq:V_3D}) will
have to be estimated, with a total of 6 parameters. In the
continuous-space case illustrated in Figure \ref
{fig:extrcoeff_climreg}, all coefficients $\alpha$ and $c$ are linearly
interpolated around the north/south border. We also considered
different mixtures of the above possible coordinates. In all cases, we
used longitude and latitude, plus possibly the elevation, region number
and mean snow depth, or combinations of these three coordinates. In
total, $65$ types of models were considered, each of them being
estimated for one Smith and nine Schlather processes, giving $650$ fits
in all. A description of the 65 model types is given in the
Supplementary Materials [\citet{BlanchetDavison2011Supp}]. All were
estimated using the iterative profiling algorithm of Section \ref
{sec:estim_practice}.

%f8 ###
\begin{figure}[t]

\includegraphics{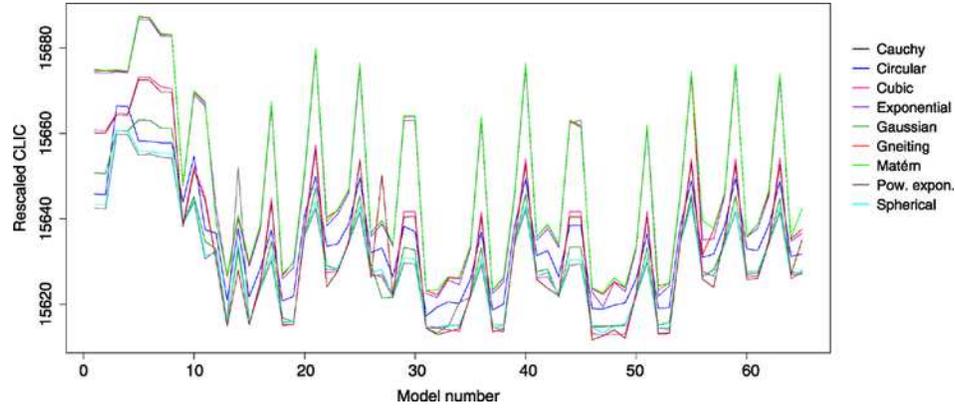}

\caption{Rescaled $\CLIC$ values for all $65\times9$ fitted Schlather
models.}\label{fig:tic}
\end{figure}

%s6.2 ###
\subsection{Model comparison}

A summary of the $\CLIC$ values for the 585 fitted Schlather models,
rescaled by division by $D-1$ in order to give log-likelihood values
that would correspond to independent data, is shown in Figure \ref
{fig:tic}. There are relatively\vadjust{\eject} small differences among them, though
the Gneiting and Gaussian covariance functions seem to perform less
well and the spherical and circular covariance functions have the $25$
best $\CLIC$ values. These covariance functions have an upper bound and
are governed by only one parameter. Schlather's model always performs
better than Smith's model, whatever the chosen covariance function: the
rescaled $\CLIC$ with Smith's model is between 30 and 300 units higher
than with Schlather's model, with a minimum value of 15,650 attained for
model 47. Whether with Smith or Schlather models, the same patterns
appear. In particular, the first eight models, which perform poorly,
correspond to models in Euclidean space, without climate space
transformation. The benefit of working in a transformed space in order
to allow for anisotropy is thus clear. This effect is particularly
striking for Smith's model, for which it is equivalent to saying that a
nonspherical $\Sigma$ matrix (see Section \ref{sec:smith_model}) should
be used: there is a difference of 300 between the lowest rescaled
$\CLIC$ values in the Euclidean and climate spaces. The models numbered $10$,
$11$, $17$, $21$, $25$, $29$, $30$, $36$, $40$, $44$, $45$, $51$, $55$,
$59$ and $63$, which are also poor, correspond to cases when neither
elevation nor the mean snow depth are considered [\citet
{BlanchetDavison2011Supp}]. As the mean snow depth is strongly related
to elevation, the latter is a very important climate coordinate. It
seems to be more informative than the mean snow depth; models using
elevation but not mean snow depth as a coordinate always have lower
$\CLIC$ values than in the converse case.

%t1 ###
\begin{table}[t]
\caption{Parameters (standard errors) of the model selected by $\CLIC$:
Schlather's model with spherical covariance function, two climate
transformations but a continuous space (the band around the north/south
border is about $5$ km wide)}\label{tab:param}
\vspace*{-2pt}
\begin{tabular*}{\tablewidth}{@{\extracolsep{\fill}}lccccc@{}}
\hline
\multicolumn{6}{@{}c@{}}{\textbf{Covariance parameter}}\\
\hline
\multicolumn{6}{@{}c@{}}{$447.45\ (43.32)$}\\[6pt]
\hline
&\multicolumn{5}{c@{}}{\textbf{Climate space parameters}}\\[-5pt]
&\multicolumn{5}{c@{}}{\hrulefill}\\
& \textbf{Main direction} & \textbf{Latitude (km)}& \textbf{Elevation (km)}& \textbf{Mean snow} & \textbf{Region num.} \\
& \textbf{(radian)}& & & \textbf{depth (cm)}& \\
\hline
North & $0.36\ (0.03)$ & $4.98\ (0.80)$ & $274.7\ (35.4)\hphantom{0}$ & $1.26\ (0.46)$
& $\times$ \\
South & $0.17\ (0.06)$ & $4.70\ (1.16)$ & $406.5\ (161.3)$ & $6.41\ (2.45)$
& $449.4\ (37.4)$ \\
\hline
\end{tabular*}
\end{table}

%s6.3 ###
\subsection{Selected model}

$\!$According to the $\CLIC$, the best fit is given by Schlather's model
with spherical covariance function, and a 5-dimensional climate space
$\mathcal X$ of coordinates (longitude, latitude, elevation, region
number, mean snow depth) with different transformations in the north
and south but imposing space continuity; this, model number 47 in
\citet{BlanchetDavison2011Supp} has a $\CLIC=15\mbox{,}611.66$. This\vadjust{\eject} means that two
$V$ matrices are estimated, each of the form (\ref{eq:V_4D}) but in
five dimensions, and thus having five parameters: the main direction of
dependence, and the four parameters $c$ associated to the latitude,
elevation, region number and mean snow depth. Since the region number
is a binary variable, the $c$ value for the northern region can be
fixed equal to zero. The range parameter of the spherical covariance
function and the width of the band between the regions are also
estimated, for a total of $11$ parameters, whose estimates and standard
errors are shown in Table \ref{tab:param}. As the pairwise likelihood
is not differentiable with respect to the band width, no standard error
is given for it. The second- and third-best fits are also obtained with
spherical covariance functions with similar models as in Table \ref
{tab:param} but without the mean coordinate (model number~49, with
$\CLIC=15\mbox{,}612.03$) or the region number coordinate (model number 46,
with $\CLIC=15\mbox{,}612.56$), that is,\ using a four-dimensional
space~$\mathcal X$. In the latter case, values of the estimated coefficients
are such that the northern and southern regions are disjoint in the
climate space, although no region number coordinate is used to separate
them. These two models perform similarly because the mean coordinate
should provide information about the local variability of snow depth,
part of which agrees with the regional division between the northern
and southern slopes; thus, the mean coordinate and region number carry
similar information. According to Figure \ref{fig:tic}, it seems better
to use both coordinates, but using one of them increases the $\CLIC$
only very slightly.

It is no surprise that in Table \ref{tab:param}, elevation is the most
influential coordinate in the climate distance, and thus in the
dependence function. In the north, for example, dependence between two
stations at the same elevation but $10$ km apart along the main
direction of dependence, an angle of $\alpha=0.36$ radians in the sense
of an Argand diagram, at the same elevation but $2$ km apart
perpendicularly to the main direction of dependence, and at the same
latitude and longitude but\vadjust{\eject} $40$ m apart in elevation, are all equal.
An interesting feature is the main direction of dependence in the
northern region, which can be explained by two facts:
\begin{enumerate}
\item due to the strong elevation effect, the north slope of the Alps
(the mountainous part of the northern region) is very weakly dependent
on the Plateau (the low-elevation part of the northern region). But
both subregions are oriented along the North Alpine ridge, and
dependence is thus higher in this direction;
\item this direction is also broadly that of the two widest valleys in
Switzerland, the Rhone and Rhine valleys, as shown by the main green
valleys in~Figu\-re~\ref{fig:map_swiss}. These are wide enough to direct
snow-bearing clouds along them, thus inducing strong directional
dependence of precipitation.
\end{enumerate}
The high value associated to the region number coordinate gives the
lowest possible dependence, $\theta_{xx'}=1.707$, between extremes in
the northern and southern regions.

%f9 ###
\begin{figure}[b]
\vspace*{-2pt}
\includegraphics{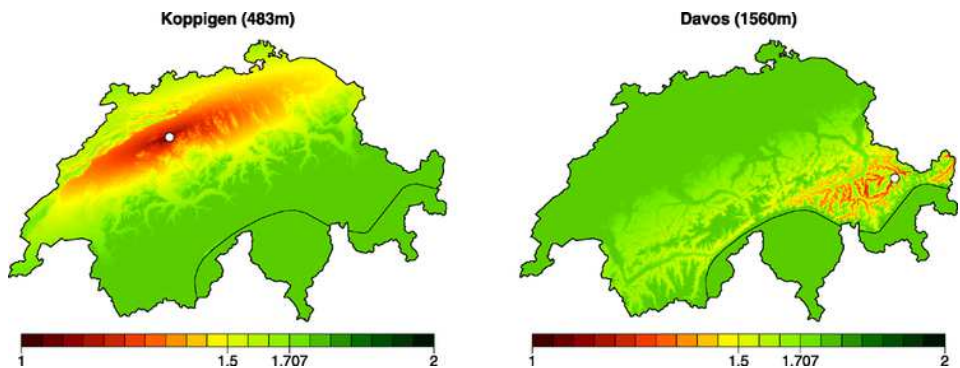}%
\vspace*{-2pt}
\caption{Pairwise extremal coefficient with Koppigen and Davos (white
circles) predicted by the selected max-stable model.}
\label{fig:map_theta_bestmod}
\end{figure}

Figure \ref{fig:map_theta_bestmod} shows maps of the estimated pairwise
dependence under the max-stable model of Table \ref{tab:param},
obtained by extrapolating the mean snow depth at ungauged stations
where no data are available. To do this, we performed spatial kriging
with a spline dependence on elevation, to allow for the fact that
temperatures at stations below $800$ m may exceed $0^\circ$C even when
it is snowing at higher altitudes, leading them to suffer rain rather
than snow. The resulting smooth mean process was successfully validated
on the additional $15$ stations [\citet{BlanchetDavison2011Supp}].
Figure \ref{fig:map_theta_bestmod} clearly shows both the elevation
effect and the weak north/south dependence. The low bandwidth, of about
$5$ km, induces an abrupt change of the extremal coefficient around the
north/south border.

\vspace*{-2pt}
%s6.4 ###
\subsection{Model checking}

\begin{figure}[t]

\includegraphics{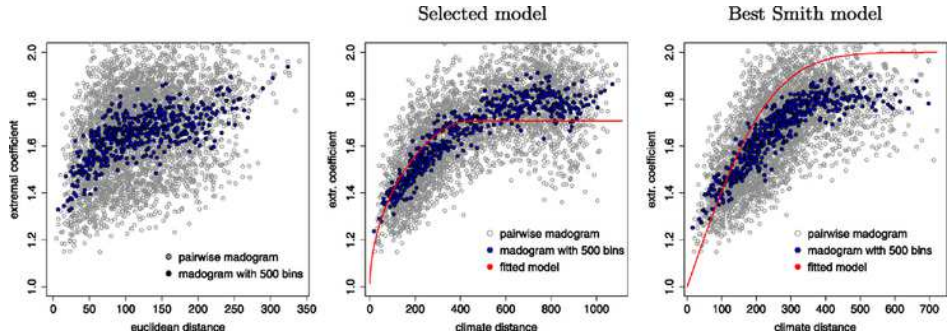}

\caption{Extremal coefficient for pairs of stations as a function of
the distance between them, in Euclidean space (left plot) or climate
space (center and right). The red curve is the extremal coefficient
curve for the corresponding max-stable model.}\label
{fig:comp_extrcoeff_bestmod}
\end{figure}

For a first check on the quality of the selected model, we compare its
predicted extremal coefficients, obtained by replacing the parameters
involved in (\ref{eq:extcoeff_schlather}) by their estimates\vadjust{\eject} from Table
\ref{tab:param}, with the naive estimators of \citet
{schlather03} or the
madogram-based estimator of \citet{cooley06b}. As the extremal
coefficients~(\ref{eq:extcoeff_schlather}) are functions of distance
between stations, we plot naive and predicted extremal coefficients
against distance. Figure \ref{fig:comp_extrcoeff_bestmod} shows such
comparisons for our selected model and for the best Smith model. For
clarity, we only show the madogram-based estimator of \citet{cooley06b},
with and without binning. The naive estimators of \citet{schlather03}
give essentially the same picture, but with slightly higher variability.

Figure \ref{fig:comp_extrcoeff_bestmod} shows that the Smith model fits
the data less well than the Schlather model. In particular, the
extremal coefficient curve of the Smith model crosses the point cloud
for the binned madogram, whereas our selected model follows it quite
well up to a climate distance of $400$, and then underestimates it. A
limit of about $1.8$ would be expected from the madogram, but cannot be
attained with Schlather's model; see Section \ref{sec:discussion}.

%f10 ###

Another way to check our model is to compare the empirical distribution
of maxima of subsets of stations, that is,\ $Z^*_{\mathcal A}=\max\{
Z^*(x_i), x_i \in\mathcal A\}$, with maxima predicted by the selected
model. The distribution of $Z^*_{\mathcal A}$ under the selected model
is known analytically only when $\mathcal A$ comprises two stations,
but samples of $Z^*_{\mathcal A}$ can be simulated for any $\mathcal
A$. Since realizations $z^*_{\mathcal A}$ of $Z^*_{\mathcal A}$ are
available for $K=43$ years, one can compare the empirical quantiles of
$Z^*_{\mathcal A}$ with the simulated ones. More precisely, given a
subset $\mathcal A$, we simula\-te~$M$ independent series ${z^{*(m)}_{\mathcal
A}}$ of length $K$, and thus obtain $M$ replicates of the
observed Fr\'{e}chet series $z^*_{\mathcal A}$. Ordered values of
observed $z^*_{\mathcal A}$ can then be compared with ordered values of
the ${z^{*(m)}_{\mathcal A}}$ as a graphical test of fit. Pointwise
and overall confidence bands can also be derived from these simulations
[\citet{davison97}, Section 4.2.4].

\begin{figure}

\includegraphics{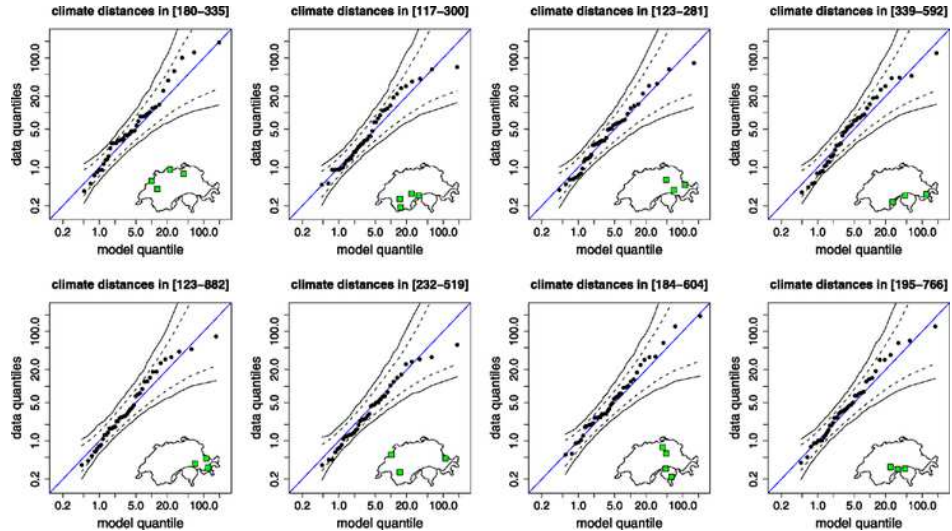}

\caption{Comparison of empirical and model quantiles for annual maxima
of groups of stations not used in the fitting. The stations used for
each panel are shown in its map, and the envelopes are 95\% pointwise
and overall confidence bands obtained from $M=5\mbox{,}000$ simulations.}\label
{fig:comp_groupstat_bestmod}
\end{figure}

Figure \ref{fig:comp_groupstat_bestmod} uses this approach to compare
fitted and empirical distributions for different groups of three or
four stations taken from the 15 not used to fit the model, some groups
being tightly clustered, and others being dispersed. The fit seems to
be broadly satisfactory in all cases. Even the dependence between
stations whose climate distance is larger than $500$ units seems to be
well-modeled, despite the mismatch between the fitted and empirical
pairwise extremal coefficients at such distances seen in Figure \ref
{fig:comp_extrcoeff_bestmod}.

%f11 ###

%s6.5 ###
\subsection{Risk analysis}

\begin{figure}

\includegraphics{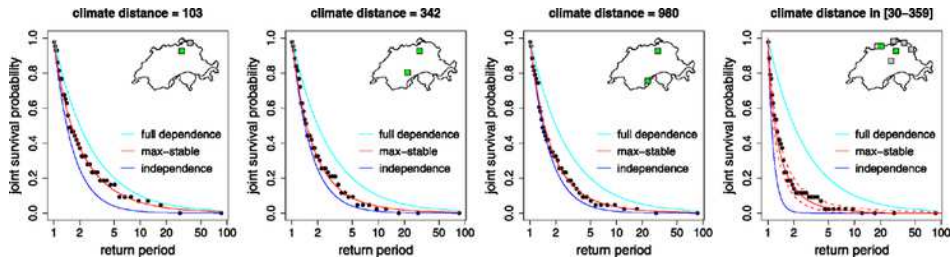}

\caption{Risk analysis of groupwise annual maxima: joint survival
probability versus return period. In the right-hand of each panel the
envelope is a 95\% pointwise confidence band obtained from $M=5\mbox{,}000$
simulations. Stations indicated in green were not used for
fitting.}\label{fig:risk_groupstat_bestmod}
\end{figure}

For risk management it is important to be able to assess how extreme
events are likely to occur in the same year in different places. A
first answer to this question can be obtained by computing
probabilities of the form $\pr[\{Z^*(x)>z, x \in\mathcal A\}]$ for a
group of stations $\mathcal A$ and different high levels $z$. Figure
\ref{fig:risk_groupstat_bestmod} plots such probabilities for different
groups~$\mathcal A$ when $z$ is the $r$-year return level of the unit
Fr\'{e}chet distribution. By back-transformation from equation (\ref
{eq:transf_frechet}), this is equivalent to computing the joint
survival distributions $\pr[\{Z(x)>\mathrm{RL}_r(x), x \in\mathcal A\}]$
where $ \mathrm{RL}_{r}(x)$ denotes the $r$-year return level at station
$x$, that is,\ the probability that all stations in~$\mathcal A$
receive more snow a given year than their $r$-year return level. Under
independence, this probability equals $r^{-|\mathcal A|}$ for any
possible set $\mathcal A$, where $|\mathcal A|$ is the number of
stations in $\mathcal A$, whereas it equals $r^{-1}$ under full
dependence. Figure \ref{fig:risk_groupstat_bestmod} shows very good
agreement between the observed and predicted distributions using the
model, whereas the risk is underestimated under the hypothesis of
independent stations and overestimated under the hypothesis of full
dependence. The underestimation is more striking for quite dependent
stations, such as those in the left-hand panel of Figure \ref
{fig:risk_groupstat_bestmod}. When distance increases, the difference
between the dependent and independent cases is less striking but our
max-stable model fits better even for pairs of stations that are $980$
climate distance units apart; this is almost the largest climate
distance between pairs of stations. The right-hand panel corresponds to
a~group of seven stations in the eastern Plateau. Our model clearly
gives more realistic risk probabilities than does the independence
assumption. Extreme snow events in the low-elevation Plateau generally
occur over a large region due to the easy weather circulation. A
typical example is the extraordinary snowfall event that occurred on
March 5th 2006 over the entire Plateau, with snow measurements of $54$
cm at Zurich, $49$ cm at Basel and $60$ cm at Sankt Gallen. This was
the largest snow depth recorded since 1931 [\citet{zanini06}].

%f12 ###

%s7 ###
\section{Discussion}\label{sec:discussion}

The models discussed here are a step toward modeling spatial dependence
of extreme snow depth. They are based on the \citet{smith91} and
\citet
{schlather02} max-stable representations, designed to model extreme
snow depth explicitly. In particular, they can account in a flexible
way for the presence of weakly dependent regions. They involve a
climate transformation that enables the modeling of directional effects
resulting from phenomena such as weather system movements. In the
proposed methodology, model fitting is performed by using a
profile-like method for maximizing the pairwise likelihood function,
and model selection is performed using an information criterion.

We applied this methodology to $86$ stations with recorded snow depth
maxima. Performance of the selected model at small and large scales was
assessed on these stations, together with $15$ other stations, by
comparing empirical and predicted distributions of group of stations.
By accounting for spatial dependence, our model gives clearly more
realistic probabilities of extreme co-occurrence than would a
nonspatial model. Such quantities are important for adequate risk management.

Considered as a whole, the max-stable models proposed in this paper
constitute a family of flexible models that could potentially be
applied to other kinds of climate data, in particular, extreme
precipitation and temperature. Further improvements could nevertheless
be investigated, as discussed below.

In this paper we focus on modeling the spatial dependence of extremes,
rather than on the marginal distributions. A first step was thus to
transform maxima from their original scale to a common unit Fr\'{e}chet
distribution. In the application to snow depth data, this
transformation was done by using the GEV distributions fitted to the
time series, considered separately. A fuller spatial model would
consider the three GEV marginal parameters as response surfaces. Using
the models presented in this paper, one could then simultaneously
estimate the spatial dependence and the spatial intensity of maxima,
following \citet{padoan10} and \citet{davison10}. These
authors use
simple functions of longitude, latitude and elevation, but the very
complex Alpine topography results in an extremely variable pattern of
snow, and we were unable to find satisfactory marginal response
surfaces for our application. \citet{blanchet10} describe other
approaches that appear to be more satisfactory, but modeling of the
margins requires more investigation. Time could be used as a covariate
in order to allow for the potential impact of climate change on extreme
snow events; for example, the retreat of the glaciers is strongly
affecting microclimates at high altitudes. This notwithstanding,
exploratory work suggests that although climate change has affected
mean snow levels [\citet{marty08}], its effect on extreme snow
events is
not yet discernible, except possibly at low elevation [\citet
{laternser03}].

A second improvement might be the consideration of event times, which
could be incorporated into the pairwise maximum likelihood procedure
[Ste\-phenson and Tawn, (\citeyear{stephenson05})]. For our data, the co-occurrence of annual maxima
is quite variable. For winters such as those of 1975 and 2006, snow
depth reached its maximum almost simultaneously all over Switzerland.
For winters such as those of 1980, 2007 and 2008, the annual maxima
occurred at quite different dates; see the Supplementary Materials,
\citet{BlanchetDavison2011Supp}. Including this information by
modifying the pairwise likelihood contribution of maxima occurring
simultaneously at two stations might yield more precise inferences, as
shown in \citet{davison10}.

Last but not least, this article has used only snow data gathered from
measurements in flat, open and not too exposed fields. Extrapolation to
steep, windy and forest terrains may thus be unsatisfactory. In
particular, preferential deposition of snow [\citet{lehning08}] may
imply that snow depth on slopes is more extreme than on representative
flat fields. This could have important implications for avalanche risk
[\citet{lehning06}]\vadjust{\eject} but could not be considered here due to lack of
data. This could be investigated using data from automatic stations
located at higher elevations, mostly above 2,200 m, and in various
terrains, though such data are unfortunately available only for about
ten years. A spatial model for exceedances over high thresholds [\citet
{davison90}] would be a valuable addition to the extreme-value toolkit
for dealing with spatially-dependent short time series.

\section*{Acknowledgments}

We thank two referees, an associate editor, the editor and the other
project participants, particularly Michael Lehning, Christoph Marty,
Simone Padoan and Mathieu Ribatet, for helpful comments. Most of the
work of Juliette Blanchet was performed at the Institute for Snow and
Avalanche Research, SLF Davos.

\begin{supplement}[id=suppA]
% \sname{Supplement A}
\stitle{Supplementary Material for ``Spatial modeling of extreme snow depth''}
\slink[doi]{10.1214/11-AOAS464SUPP}
\slink[url]{http://lib.stat.cmu.edu/aoas/464/supplement.pdf}
\sdatatype{.pdf}
\sdescription{This contains example time series of data, and further
discussion of the estimation algorithm and of the fitted models.}
\end{supplement}

\printaddresses

\end{document}